\def\year{2022}\relax
\documentclass[letterpaper]{article} 
\usepackage{aaai22}  
\usepackage{times}  
\usepackage{helvet}  
\usepackage{courier}  
\usepackage[hyphens]{url}  
\usepackage{graphicx} 
\usepackage{natbib}  
\usepackage{caption} 
\DeclareCaptionStyle{ruled}{labelfont=normalfont,labelsep=colon,strut=off} 
\usepackage{algorithm}
\usepackage{algorithmic}

\usepackage{xcolor}
\usepackage{xspace}
\usepackage{graphicx}
\usepackage{comment}
\usepackage{multirow}
\usepackage{makecell}

\usepackage{subcaption}
\newcommand{\yiftach}[1]{{\textcolor{brown}{[Yiftach: #1]}}}

\newcommand{\MethodName}{OWSD\xspace}
\usepackage{newfloat}
\usepackage{listings}
\floatstyle{ruled}
\newfloat{listing}{tb}{lst}{}
\floatname{listing}{Listing}
\title{Secure Machine Learning in the Cloud Using One Way Scrambling by Deconvolution}

 \author{Yiftach Savransky,\textsuperscript{1}
Roni Mateless,\textsuperscript{1}
Gilad Katz\textsuperscript{1}\\
\textsuperscript{1}{Ben-Gurion University of the Negev}\\
yiftachs@post.bgu.ac.il,
mateless@post.bgu.ac.il, 
giladkz@bgu.ac.il}

\begin{document}

\maketitle

\begin{abstract}
Cloud-based machine learning services (CMLS) enable organizations to take advantage of advanced models that are pre-trained on large quantities of data. The main shortcoming of using these services, however, is the difficulty of keeping the transmitted data private and secure. Asymmetric encryption requires the data to be decrypted in the cloud, while Homomorphic encryption is often too slow and difficult to implement. We propose One Way Scrambling by Deconvolution (OWSD), a deconvolution-based scrambling framework that offers the advantages of Homomorphic encryption at a fraction of the computational overhead. Extensive evaluation on multiple image datasets demonstrates OWSD's ability to achieve near-perfect classification performance when the output vector of the CMLS is sufficiently large. Additionally, we provide empirical analysis of the robustness of our approach. 
\end{abstract}

\section{Introduction}

Recent advances in Cloud-based Machine Learning Services (CMLS) capabilities have made it possible for individuals to access state-of-the-art algorithms that until recently were only accessible to few. 
These capabilities, however, come with two major risks: \textit{a)} the leakage of sensitive data as it's being sent to and/or processed by the cloud, and; \textit{b)} the ability of third parties to gain insight into the organization's data by analyzing the output of the ML model in the cloud.

Given that organizations must transmit their data to the cloud to use cloud services, the former are faced with three options: \textit{a)} Send the data in encrypted form to the cloud, where it will be decrypted, analyzed and then sent back in encrypted form (e.g., using asymmetric encryption); \textit{b)} Employ Homomorphic Encryption (HE) techniques that enable the cloud to process the data in its encrypted form; and, \textit{c)} apply differential privacy techniques which introduce varying degrees of noise to the data. Each of these approaches, however, has shortcomings.

The first option -- decryption in the cloud -- enables the cloud provider unfettered access to the organization's private data, and is unacceptable in many cases. HE keeps the data private, even when it is being processed in the cloud, but it is slow and puts various limitations on the operations that can be performed so as not to prevent the decryption at the end of the process. Other encryption solutions that offer some or all of the capabilities of HE also exist, but they too require cooperation from the service provider (i.e., the cloud-based service) and often have other prerequisites such as non-collusion among participating servers or dedicated hardware. Finally, differential privacy techniques do prevent attackers from inferring specific details from the data, but they do not prevent obtaining a general sense of the processed data and they come with the price of reduced performance.

In this study we propose One Way Scrambling by Deconvolution (\MethodName), an approach that provides the main advantage of Homomorphic encryption---not sharing the unencrypted data with any party, including the CMLS---but with a computational overhead that is smaller by orders of magnitude. Our approach uses a randomly-initialized deconvolutional architecture to scramble the input, which is then sent to the CMLS. The CMLS processes the input in the same manner as it would any unencrypted data, and produces a classification vector. Finally, a small ``translator'' neural architecture within the organization receives the classification vector produced by the cloud and combines it with an embedding of the original image to predict the classification the cloud service would have given to the original image.

\MethodName has several important advantages compared to existing methods. First, the use of randomly-initialized deconvolutional networks creates scrambled images that are both incomprehensible to human observers and whose classification in the cloud does not disclose their original label. Secondly, as the deconvolutional network is randomly initialized and not trained, replacing the ``key'' (i.e., network weights values) of our scrambling mechanism is computationally inexpensive, as we only need to retrain our translator architecture. Moreover, multiple scrambling keys can be used simultaneously, thus both improving \MethodName's performance and making decryption efforts more challenging. Thirdly, our approach does not require any coordination with the CMLS.

Our contributions in this study are as follows:
\begin{itemize}
	\item We present \MethodName, a scrambling-based approach that has many of the advantages of HE at a fraction of the computational cost.
	\item We perform extensive evaluation and show that our approach can reach up to 99.9\% of the performance of the CMLS when the number of labels in the confidential data is smaller than that of the cloud. This is also the case when the labels of the confidential data are not in the cloud's training data (i.e., transfer learning).
	\item We perform an extensive empirical analysis of our approach and provide an upper bound on the number of images that can be securely scrambled by any one key of our approach.
	
\end{itemize}

\section{Related Work}

Existing solutions can be roughly divided into two groups: encryption-based and differential privacy-based.

\subsection{Encryption-based solutions}
In this study we focus on approaches that enable the application of privacy-preserving Machine Learning (ML) in the cloud. These solutions can also be divided into two groups: one in which the cloud-based service provider has access to the plaintext, and one in which it does not.

The former group of solutions utilizes common public/private key settings (e.g., RSA \cite{milanov2009rsa}) and is in widespread use. In Li, Ping, et al. \cite{li2018privacy},  the authors present a multi-step framework for the exchange of public keys. A semi-honest model -- a scenario where the cloud provider is not compromised -- is also assumed. Similar approaches have also been proposed by \cite{li2018privacy,li2018privacymulti}.

The latter group of solutions, where the cloud provider does not have access to the plaintext, can be divided into hardware and software-based solutions. In hardware-based solutions, multiple studies \cite{hunt2018chiron,tramer2018slalom,hynes2018efficient} utilize the SGX architecture which possess a secure enclave for the processing of sensitive information. In Chiron \cite{hunt2018chiron}, for example, the authors demonstrate how SGX could be effectively used to run deep neural architectures. This approach, while effective, requires the installation of a dedicated client both by the client and the service provider, as well as adaptation of the ML model, and are costly and difficult to implement.

Software-based encryption solutions that without the plaintext from the CMLS, though diverse, mostly revolve around the use of HE. In \cite{li2017multi} the authors propose two Fully Homomorphic Encryption (FHE) schemes for privacy-preserving deep learning. In \cite{hesamifard2018privacy} and \cite{hesamifard2017privacy}, the authors demonstrated that it is feasible to train neural networks using encrypted data, while in \cite{gilad2016cryptonets} the authors detailed a method to convert learned neural networks to neural networks that can be applied to encrypted data (using leveled HE). While they enable full protection of one's data, Homomorphic solutions are slow, require large amount of memory and require prior knowledge of the type of queries, so that the appropriate HE scheme can be selected to encrypt the raw data \cite{li2017multi}.

\subsection{Differential privacy-based solutions}

Differential privacy is a general term for techniques that obfuscate the data (i.e., inject noise) in order to ensure that the details of individual records cannot be inferred \cite{dwork2008differential}. Generally, these techniques sacrifice some accuracy in the interest of privacy. The main drawback of this approach is that there is still a way of inferring the general type and characteristics of the data.

In \cite{bonawitz2017practical} the authors use aggregations of data collected from mobile devices, which are then sent to a CMLS. A similar approach is presented in \cite{yang2018machine,osia2020hybrid}, who also incorporate the injection of noise into the data. A different approach is presented in \cite{wang2018not}, where a deep architecture is divided into two parts: the smaller (initial part) is installed locally on the client, and the bulk of the architecture is in the cloud. The data passes through the client (compressed and encrypted) and then sent to the cloud. Noise is also added to ensure differential privacy. While the latter solution bears some similarity to our proposed approach, the two networks have to be trained jointly and cooperation between the cloud and the client is necessary. Moreover, the cloud-based provider has access to the final classifications and has some ability to reconstruct the data---a shortcoming not shared by \MethodName.

\section{The Proposed Method}
\label{sec:proposedMethod}

\begin{figure}[h!]
	\centering
	\includegraphics[width=0.9\columnwidth]{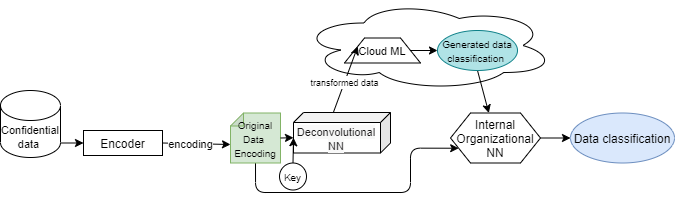}
	\caption{\MethodName's inference process.}
	\label{fig:encrypArchitecture}
\end{figure}

Our proposed approach is presented in Figure \ref{fig:encrypArchitecture}. \MethodName consists of three components: \textit{a)} an \textit{encoder}, that creates an embedding of the original image; \textit{b)} a \textit{generative model}, that receives the embedding as input and creates the encrypted image, and; \textit{c)} an \textit{internal inference network}, which receives the classification vector for the encrypted image from the cloud and infers the labels of the original (unencrypted image). We now review these components.\\

\noindent \textbf{The Encoder.} The goal of this component is to create a meaningful condensed representation of the original (unencrypted) image. Multiple convolutional architectures would be suitable for this task. In our experiments we used several pre-trained and well-known image classification architectures (e.g., ResNet50, ResNet50V2, ResNet101) whose softmax layers were removed to create our embeddings.\\

\noindent \textbf{The Generative Model.} The goal of this component is to create a scrambled version of the original image. We use a deconvolutional architecture that receives the embedding generated by the encoder and outputs a new image. The new image is scrambled because it is generated by a \textit{network whose weights are randomly initialized}. This set of randomly-initialized weights serves as \MethodName's encryption key. Next, the newly-generated (i.e., scrambled) image is sent to the CMLS. The service processes the image as it would any unencrypted input, producing a classification vector. This vector is then sent back to the secure organizational network. It is important to note that \textit{only scrambled images ever leave the organizational network}, both during the training of model (on which we elaborate later in this section) and when it is applied on new data. \\

\noindent \textbf{The Internal Inference Network (IIN).} The goal of this component is to infer the label of the original (unscrambled) image. This dense neural network receives as input the embedding of the original image (produced by the Encoder) and the classification vector produced by the CMLS for the encrypted image. The network then \textit{infers the true label of the original image}. This network can be easily trained using a small number training inputs---training required on average 4-5 minutes on a standard laptop.

\subsection{The Training Phase}
The only component of \MethodName that requires training is the IIN. To train this component in a fully secure manner, the organization requires a small dataset of images $D$ whose real labels are known. These images can then be scrambled, sent to the cloud, and have their classification vectors retrieved. We then use the embedding of each image (the output of the Encoder) and the CMLS's classification vector to train the IIN to predict the true label of the original image.

We consider the requirement that organizations have a small labeled dataset to be reasonable, since most organizations have proprietary knowledge they wish to protect and therefore able to leverage for \MethodName's training. If such a dataset is not available, however, it is also possible to use publicly-available images to train the inference network. We explore such a scenario in use-case 3 of Section \ref{subsec:evaluationResults}. 

\subsection{Why Does Our Approach Work?}
The main insight behind \MethodName's encryption process is that convolutional (and deconvolutional) architectures identify and transform latent patterns in the data. These patterns are very difficult to reverse-engineer because of the non-linear nature of the neural net's activation function and the large number of operations, but they are maintained nonetheless. These transformations are performed in a consistent (yet complex) manner, and will therefore elicit consistent behavior from the CMLS (i.e., a classification vector).

Our ability to translate the classification vector of the encrypted image into a label for the original image is not without foundations in existing literature. Our work builds upon two areas of research: membership inference attacks \cite{shokri2017membership} and distillation \cite{hinton2015distilling}. In membership inference attacks, an attacker can learn which samples were used to train a black-box model by analyzing its output vector. This field of research shows that one can derive meaningful insights regarding the neural network's inner workings from its output vector. In distillation, a single neural network is used to learn the decision process of an entire ensemble, and is thus able to achieve similar performance at a fraction of the computational resources. The ideas behind these two approaches are combined in \MethodName: we analyze the output vectors for encrypted images, and train a small neural network to distill the performance of the cloud-based model, although for predicting possibly different labels than the cloud's.

\subsection{Why is our approach secure?}
One-time pad encryption (OPE) is known to be unbreakable, as long as \textit{a)} it is of the same length as the encrypted text, and; \textit{b)} it is used only once \cite{matt2013one}. Our approach builds upon OPE, but with several modifications: first, we use an `encryption pad' (i.e., the deconvolutional network) that is larger by orders of magnitude than the encrypted text. This makes decryption difficult even when the pad is used for scrambling multiple samples. Secondly, the deconvolutional network's use of the ReLU activation function means that negative values generated throughout the network will be transformed to zero. This trait of our network adds another layer of complexity for an attacker, because multiple parameter configurations could be used to generate an input/output pair. Our analysis shows that on average 59\% of the neurons of our deconvolutional network output zeros for any input they receive. Finally, we present a comprehensive empirical analysis of the strength of our approach in Section \ref{subsec:AnalyzingRobustness}, and provide a bound on the number of images that can be safely scrambled by any given key.

\section{Experiments}

\subsection{Experimental setup}
\label{subsec:ExperimentalSetup}

\begin{itemize}
	\item We used four datasets in our experiments: \textit{a)} two variants of ImageNet \cite{deng2009imagenet}: ILSVRC2010 and ILSVRC2012; \textit{b)} CIFAR-100, and; \textit{c)} CIFAR-10 \cite{krizhevsky2009learning}.
	
	\item We used the following pre-trained architectures as our CMLS: InceptionV3 \cite{szegedy2016rethinking}, Xception \cite{chollet2017xception}, and VGG16 \cite{simonyan2014very}. The two former architecture were trained on the training set of ILSVRC2012 (i.e., ImageNet), and the latter both on CIFAR10 and CIFAR100.
	
	
	\item For the Encoder component, we used pre-trained architectures of ResNet50 and ResNet101 \cite{he2016deep}, and ResNet50V2 \cite{rahimzadeh2020modified}, that were trained on the training set of ILSVRC2012. The output of this component is an embedding vector with 2,048 entries. Please note that the same pre-trained architecture is \textit{never used simultaneously in the Encoder and as the CMLS in our experiments}.
	
	\item For our Generative model, we used the DCGAN's \cite{radford2015unsupervised} generator. The dimensions of the output image were $256x256x3$.
	
	\item Our IIN was a dense neural network with a single hidden layer, followed by a softmax layer. The input vector size was $2048 + |V|$, where $V$ is the classification vector of the cloud. We applied batch normalization and dropout, and used ADAM optimization. We used a learning rate of 0.0001 with exponential decay, and trained the network up to 40 epochs, with early stopping.

	
	\item We evaluate our approach using two metrics---top-1 and top-5 accuracy---which are the standard in image classification tasks. We calculate these metrics with respect both to the ground truth (i.e., true labels) and to the cloud's classification of the unencrypted image (i.e., we measure our ability to infer the cloud's classifications).
	
	\item All experiments in this section (i.e., training the INN) were conducted on a laptop with 16GB RAM and an Intel CPU (Intel(R) Core(TM) i7-8565U).
	
	
\end{itemize}

\subsection{Evaluation Results}
\label{subsec:evaluationResults}

We evaluate five use-cases: \textit{a)} \textit{same labels in confidential data and cloud} -- in this use-case, the cloud was trained on images with the same labels as confidential data (no overlap between the two sets of images); \textit{b)} \textit{subset of cloud labels} -- the confidential data consists of a subset of the labels on which the cloud was trained -- the goal of these experiments is to assess the impact of the relative number of labels on performance; \textit{c)} \textit{different labels in confidential data and cloud} -- these experiments evaluate our ability to perform transfer learning to new labels; \textit{d)} \textit{ensemble} -- we explore the effects of using several Encoders; \textit{e)} \textit{Varying IIN training set sizes} -- we analyze \MethodName's performance as a function of the IIN's training set size.\\


\noindent \textbf{Use-Case 1: Same Labels in Confidential Data and Cloud.} We conducted three sets of experiment, for ImageNet, CIFAR100, and CIFAR10. Table \ref{tab:Use Case 1 Results} presents the pre-trained models used as the CMLS and the Encoder component in each experiment (see Section \ref{subsec:ExperimentalSetup} for specifics on the training of each architecture). For the ImageNet experiments, we used 90\% of the validation set---45,000 images---to train the INN, and used the remaining 10\% (5,000 images) as our test set. For CIFAR10/100, we used 90\% (9,000 images) of the test set to train the INN, and the remaining 10\% for our evaluation (class ratios were maintained). We performed five experiments for each dataset, with random 90\%/10\% splits, and report the averages of all runs.

The results of our evaluation are presented in Table \ref{tab:Use Case 1 Results}. We present the INN's performance with respect to the ground truth and to the CMLS's classifications of the unencrypted images. To place our results in context, we also present the performance of the cloud model on the unencrypted images. these results are, in effect, the upper bound for the performance of our approach. The results clearly show that \MethodName is capable of inferring both the ground truth labels and the cloud's classifications. There is, however, a decrease in performance. For the prediction of true labels, for example, our accuracy was reduced by 6.2\%-9.8\% for ImageNet, by 1.6\%-5.2\% for CIFAR100, and by 0.6\%-6.8\% for CIFAR10. 

An important observation is that while there is a decrease in performance, it is smaller when the \textit{number of labels in the confidential data is small}. The reason for this phenomenon is simple: as the ratio of labels in the confidential data and that of the cloud $\frac{|L_{conf}|}{|L_{cloud}|}$ becomes smaller, each confidential data label can be expressed in a more nuanced way. This additional ``bandwidth'' is helpful because, as we show in Section \ref{subsec:analyzingCloudOutput}, the use of our scrambling approach results in much higher entropy in the CMLS's classification vector. Our next use-case therefore focuses on further analysis of the effects of $\frac{|L_{conf}|}{|L_{cloud}|}$.\\

\begin{table*}[t]
	\centering
	\begin{tabular}{|c|c|c|c|c|c|c|c|c|}
		\hline
		\multirow{3}{*}{Dataset} &
		\multirow{3}{*}{\makecell{Cloud\\ Architecture}} &
		\multirow{3}{*}{\makecell{Encoder\\ Architecture}} &
		\multicolumn{4}{c|}{Internal Inference}&
		\multicolumn{2}{c|}{Cloud Performance} \\
		& & & \multicolumn{2}{c|}{Real Labels} &
		\multicolumn{2}{c|}{Cloud Class.} &
		\multicolumn{2}{c|}{Un-encrypted images} \\
		& & & Top-1 & Top-5 & Top-1 & Top-5 & Top-1 & Top-5 \\
		\hline
		ImageNet&Xception&ResNet101& \makecell{67.9\% \\(0.007)} & \makecell{87.6\% \\(0.006)}& \makecell{69.3\% \\(0.008)}& \makecell{88.8\% \\(0.004)}& \makecell{77.7\% \\(0.005)}& \makecell{93.8\% \\(0.004)}\\
		\hline
		CIFAR100&VGG16&ResNet50V2& \makecell{64.3\% \\(0.013)}& \makecell{88.5\% \\(0.015)}& \makecell{58.1\% \\(0.004)}& \makecell{81.6\% \\(0.015)}& \makecell{69.5\% \\(0.011)}& \makecell{90.1\% \\(0.004)}\\
		\hline
		CIFAR10&VGG16&ResNet50V2& \makecell{86.5\% \\(0.017)}& \makecell{99.2\% \\(0.004)}& \makecell{84.8\% \\(0.014)}& \makecell{98.8\% \\(0.004)}& \makecell{93.3\% \\(0.009)}& \makecell{99.8\% \\(0.002)}\\
		\hline
	\end{tabular}
	\caption{Use Case 1 Results }
	\label{tab:Use Case 1 Results}
\end{table*}

\noindent \textbf{Use-Case 2: Using a subset of confidential data labels.} 
Based on use-case 1, we now revisit our ImageNet experiments (in which \MethodName had the largest gap with the original cloud performance). We used the same settings as in use-case 1, but the confidential data contained only a subset of 10/100 labels instead of the original 1,000. For each of these two settings---10 and 100---we performed five experiments, with the labels of each randomly selected. We used a 70\%/30\% train/test split with 50 images per label, resulting in 350/3,500 images for the training of the IIN and 150/1,500 images for evaluation. 

The results of our evaluation are presented in Table \ref{tab:Use Case 2 Results}. It is clear that our performance on ImageNet improves as the number of labels decreases, with \MethodName's top-1 accuracy rising from 67.9\% to 95.5\%, and its top-5 accuracy rising from 87.6\% to 99.9\%. The improvement in performance shows that we were correct in our hypothesis that  $\frac{|L_{conf}|}{|L_{cloud}|}$ is the main factor in determining our method's performance.\\

\begin{table}
	\begin{tabular}{|c|c|c|}
		\hline
		Number of labels & Top-1 & Top-5 \\
		\hline
		1000 & 67.9\% (0.007)& 87.6\% (0.006)\\
		\hline
		100 & 87.2\% (0.009)& 96.7\% (0.006)\\
		\hline
		10 & 95.5\% (0.019)& 99.9\% (0.003)\\
		\hline
	\end{tabular}
	\caption{Use Case 2: IIN's accuracy compared to ``ground truth''}
	\label{tab:Use Case 2 Results}
\end{table}

\noindent \textbf{Use-Case 3: Different labels for confidential data and cloud.} 
We now evaluate \MethodName's ability to infer labels that did not appear in the CMLS's training set. This use-case is important because it assesses whether our proposed approach is generic and transferable. For these experiments we used VGG16, which was trained on CIFAR 100 as our CMLS, and ResNet101, which was trained on ImageNet, as our Encoder. 

To generate our confidential dataset, we first identified all ImageNet labels that \textit{did not} appear in CIFAR100. Next we randomly sampled 10 labels from these labels---150 images per class, 1,500 overall. We use 70\% of these images (1,050) to train our Internal Inference agent and use the remaining 30\% for evaluation. This process is repeated five times, and the averaged results are presented in Table \ref{tab:Use Case 3 Results}. For this setting we can only present the results with respect to the original image labels, which clearly show that we are able to reach the same level of performance for images with previously-unseen labels as we do for those on which the CMLS was trained. Additionally, we once again show that our approach is most effective when $\frac{|L_{conf}|}{|L_{cloud}|}$ is small.\\


\begin{table}
	\begin{tabular}{|c|c|}
		\hline
		Metric & Internal Inference to Real Labels \\
		\hline
		Top-1 & 96\% (0.015)\\
		\hline
		Top-5 & 99.9\% (0.002)\\
		\hline
	\end{tabular}
	\caption{Use Case 3: results on ``ground truth''}
	\label{tab:Use Case 3 Results}
\end{table}

\noindent \textbf{Use-Case 4: Using an ensemble of Encoders.} The goal of this use-case is to examine whether an ensemble of Encoders can improve the performance of our proposed approach. To this end, we use multiple pre-trained networks as our Encoders. The resulting encodings are scrambled using the same Generative Model and sent to the CMLS. The classification vectors produced for each individual encoding are then concatenated and fed as a single input the to IIN.

The results of our analysis are presented in Table \ref{tab:Use Case 4 Results}. We build upon the experiments presented in Table \ref{tab:Use Case 1 Results}, and therefore the results for a single encoder are identical. When using two encoders, the network we added was ResNet50. For our third encoder we used ResNet50v2 for the ImageNet dataset and ResNet101 for CIFAR100/10. The results show that adding encoders indeed improves \MethodName's performance, although the improvement appears to be greater for the CIFAR10/CIFAR100 datasets (which concurs with our conclusion in use case 2). Using the paired-t test, we were able to determine that the two-Encoders setting significantly outperforms the single-Encoder setting with $p<0.001$, and that three-Encoders outperform the two-Encoders with a smaller but still highly significant value of $p<0.01$. \\



\begin{table*}[t]
	\centering
	\scriptsize
	\begin{tabular}{|c|c|c|c|c|c|c|c|c|c|c|c|c|c|c|c||c|c|}
		\hline
		\multirow{3}{*}{} &
		\multicolumn{12}{c|}{Internal Inference}&
		\multicolumn{2}{c|}{Cloud Performance} \\
		& \multicolumn{6}{c|}{Real Labels} &
		\multicolumn{6}{c|}{Cloud Class.} &
		\multicolumn{2}{c|}{Un-encrypted images} \\
		& \multicolumn{3}{c|}{Top-1} & \multicolumn{3}{c|}{Top-5} & \multicolumn{3}{c|}{Top-1} & \multicolumn{3}{c|}{Top-5} & Top-1 & Top-5 \\
		\makecell{datasets/ \\ \#encoders}& 1 & 2 & 3 & 1 & 2 & 3 & 1 & 2 & 3 & 1 & 2 & 3 &  &  \\
		\hline
		ImageNet& \makecell{67.9\% \\ (0.007)} & \makecell{68.6\% \\ (0.002)} & \makecell{69.5\% \\ (0.006)} & \makecell{87.6\% \\ (0.006)} & \makecell{88.9\% \\ (0.004)} & \makecell{89.1\% \\ (0.006)} & \makecell{69.3\% \\ (0.008)} & \makecell{69.9\% \\ (0.004)} & \makecell{71.4\% \\ (0.009)} &  \makecell{88.8\% \\ (0.004)} & \makecell{90.1\% \\ (0.003)} & \makecell{90.7\% \\ (0.005)} & \makecell{77.7\% \\ (0.005)} & \makecell{93.8\% \\ (0.004)}\\
		\hline
		CIFAR100& \makecell{64.3\% \\ (0.013)} & \makecell{67.2\% \\ (0.01)} & \makecell{68.4\% \\ (0.007)} & \makecell{88.5\% \\ (0.015)} & \makecell{89.8\% \\ (0.004)} & \makecell{90.4\% \\ (0.009)} & \makecell{58.1\% \\ (0.004)} & \makecell{61\% \\ (0.01)} & \makecell{62.4\% \\ (0.009)} & \makecell{81.6\% \\ (0.015)} & \makecell{84.5\% \\ (0.011)} & \makecell{85.7\% \\ (0.015)} & \makecell{69.5\% \\ (0.011)} & \makecell{90.1\% \\ (0.004)}\\
		\hline
		CIFAR10& \makecell{86.5\% \\(0.017)} & \makecell{88.1\% \\ (0.014)} & \makecell{88.9\% \\ (0.009)} & \makecell{99.2\% \\(0.004)} & \makecell{99.3\% \\ (0.004)} & \makecell{99.4\% \\ (0.001)} & \makecell{84.8\% \\ (0.0014)} & \makecell{86.2\% \\(0.015)} & \makecell{87.4\% \\ (0.008)} & \makecell{98.8\% \\(0.004)} & \makecell{99.2\% \\(0.001)} & \makecell{99.1\% \\(0.002)} & \makecell{93.3\% \\(0.009)} & \makecell{99.8\% \\(0.002)}\\
		\hline
	\end{tabular}
	\caption{Use Case 4 Results -- the effect of using multiple Encoders on performance.}
	\label{tab:Use Case 4 Results}
\end{table*}


\noindent \textbf{Use-Case 5: The Effect of the IIN's training size on performance.}
We now analyze the effect of the training set size on the performance of the IIN. While it likely that larger training sets will lead to higher accuracy, our goal is to quantify this impact. We conducted our analysis as follows: We used ILSVRC2012's validation set, from which we randomly sampled 100 labels, to train our IIN and evaluate its performance. We used InceptionV3 as our CMLS (this model was trained on ILSVRC2012's \textit{training set}), and ResNet50 as our Encoder. 

\begin{figure}[h!]
	\centering
	\includegraphics[width=0.9\columnwidth]{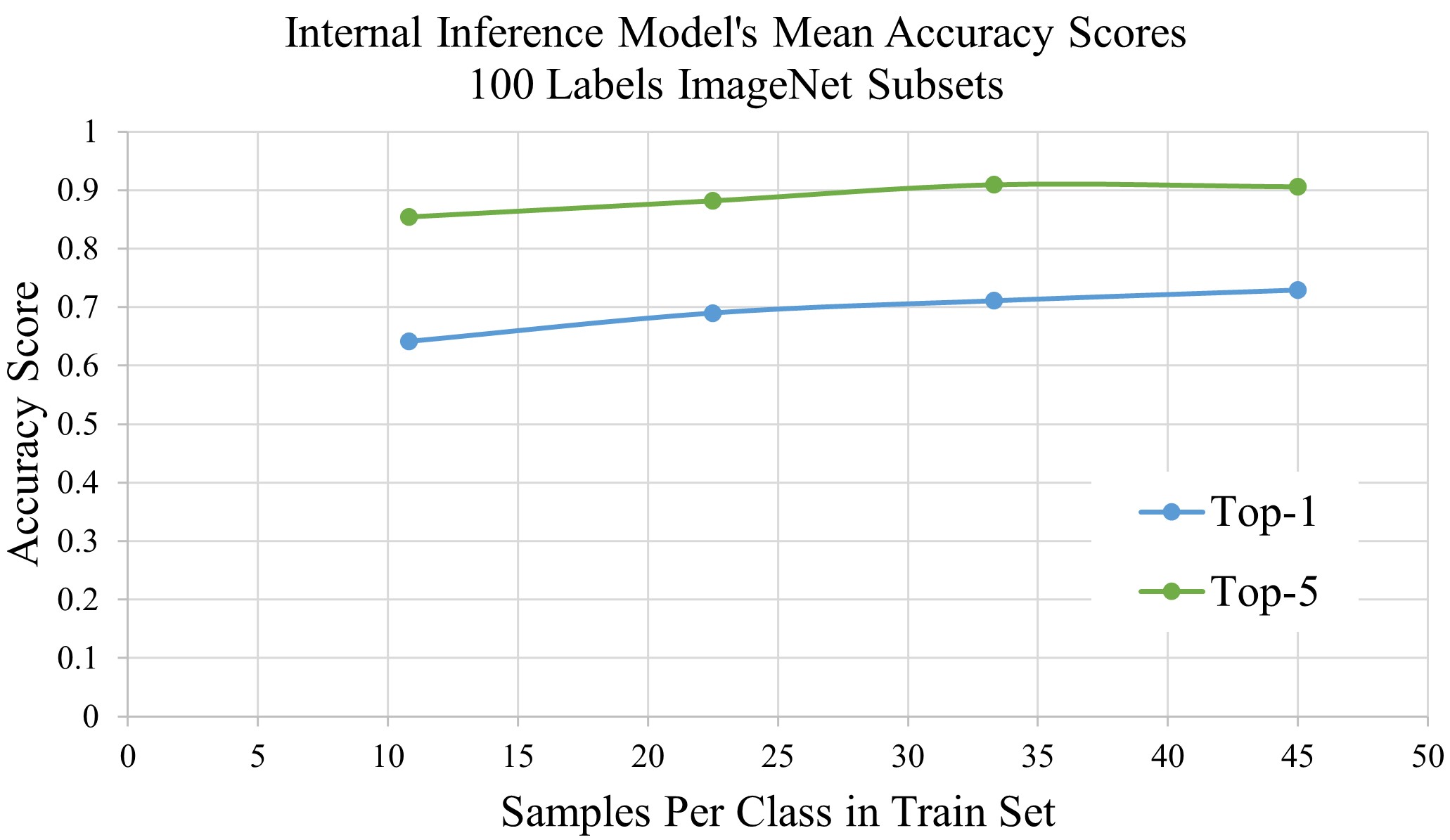}
	\caption{The top-1 and top-5 accuracy of OWSD on a randomly-sampled subset of ImageNet, consisting of 100 labels. The results are presented as a function of the number of images per label that were used in OWSD’s training set.}
	\label{fig:accScore100}
\end{figure}

We sampled different numbers of images from each of our selected 100 labels, and used them to train the IIN. We repeat this experiment five times and report the average results, which are presented in Figure \ref{fig:accScore100}. It is clear that an increase in the training set significantly improves the performance of our approach, with top-1 accuracy improving by 8.7\% and top-5 accuracy improving by 5.2\%. It is also clear that the improvement in \MethodName's performance begins to plateau at around 35 images per label, which achieved an accuracy of 90.6\% compared to the 91.2\% for 45 images, indicating that a relatively small number of training samples is needed for maximal performance.

The small number of samples that is needed to train our INN is a crucial aspect of our proposed approach. Given that each key (i.e., the randomly-initialized Generator weights) can only be used a limited number of times before an adversary can reconstruct the original images (see Section \ref{subsec:AnalyzingRobustness} for analysis), requiring a limited number of images to train the INN means that each key can be used for a longer period of time.


\section{Confidentiality Analysis}

In this section we analyze four aspects of our proposed approach and demonstrate the difficulties an attacker would face when attempting to recover the original images. First, in Section \ref{subsec:originalImage} we show that our scrambled images do not contain any information that can be understood by humans. Secondly, in Section \ref{subsec:analyzingCloudOutput} we analyze the outputs (i.e., classification vectors) produced by the CMLS for our images, and show that the former's entropy is much higher than that of unscrambled images and that the ``true'' label of the image (i.e., the ground truth) is almost never in the top-5 chosen labels. In Section \ref{subsec:scramblingReconstruct} we analyze our scrambled images and show that they are more difficult to reconstruct than their plaintext counterparts. Finally, in Section \ref{subsec:AnalyzingRobustness} we provide an empirical loose upper bound on the number of images that can be scrambled by \MethodName using a single key.


\subsection{Comparing Original and Scrambled Images}
\label{subsec:originalImage}

\begin{figure}[h!]
	\centering
	\includegraphics[width=\columnwidth]{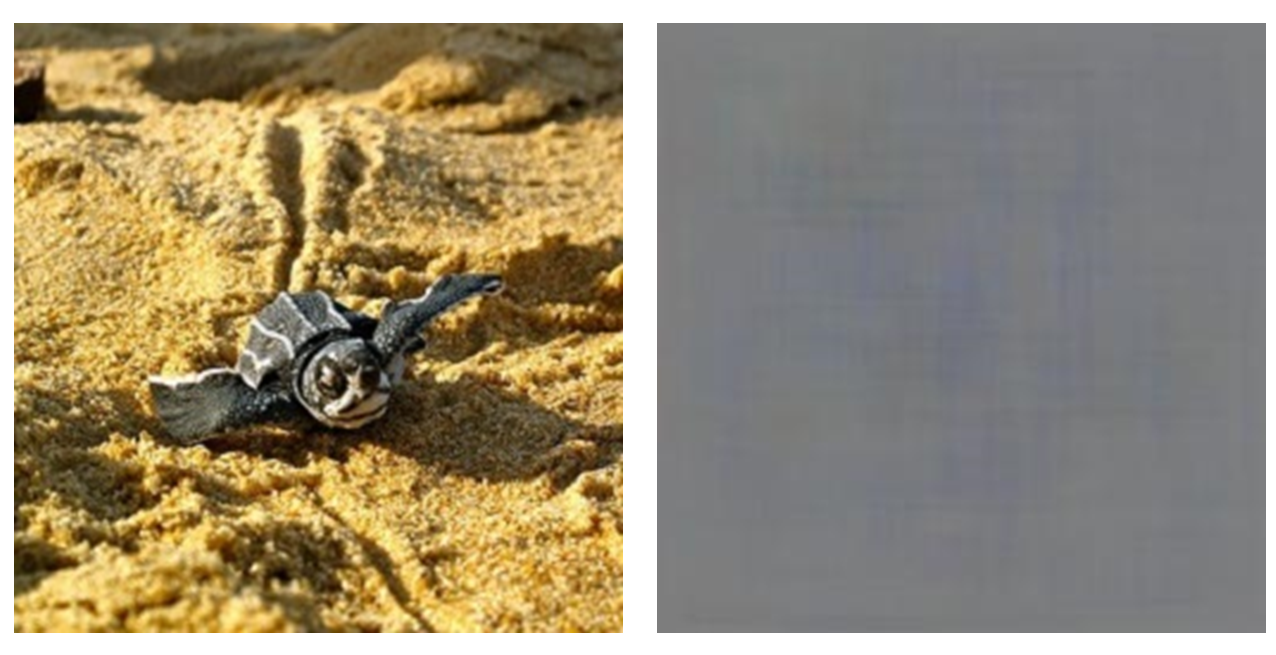}
	\caption{left) The original image. right) The same image, scrambled by \MethodName using embeddings by ResNet50.}
	\label{fig:encrypted-pair-example}
\end{figure}

We begin by addressing a simple question: can humans identify objects in our scrambled images. Figure \ref{fig:encrypted-pair-example} presents an original image and its scrambled counterpart. It is clear that no information in the scrambled image is discernible to humans using the naked eye. Similar grayscale images were produced for all the images we manually examined.

\subsection{Analyzing the Cloud Output}
\label{subsec:analyzingCloudOutput}

Our goal is to determine whether our approach effectively obfuscates the labels of scrambled images. We used the InceptionV3 and ResNet50 architectures as our CMLS and Encoder models, respectively. The training and evaluation image sets were those of use-case 1 in Section \ref{subsec:evaluationResults}.

We begin by comparing the classification vectors produced by the CMLS for each image and its scrambled counterpart. Our results, presented in Table \ref{tab:predictions-vectors-results}, show that the probabilities of any intersection of labels in the top-1 and top-5 labels is 0.07\% and 0.52\%, respectively. Moreover, the confidence scores assigned for each of the original top-1/5 labels have been reduced by an average of 76\%-87\% respectively. Finally, an analysis of the entropy of the classification vectors of plaintext and confidential images shows that the entropy of the latter is more than twice that of the former: 4.3 to 1.97. These results indicate that not only are our scrambled images not discernible to humans, but the CMLS's classification vector does not enable an easy inference of the original label(s) assigned to them.

\begin{figure}[h!]
	\centering
	\includegraphics[width=0.9\columnwidth]{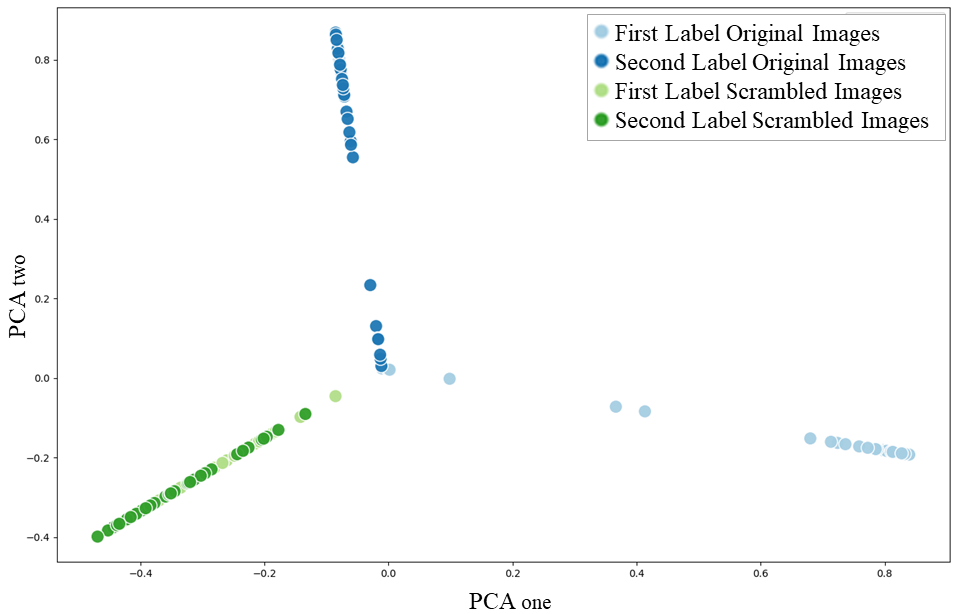}
	\caption{The classification vectors of plaintext and scrambled images of two labels, presented in a 2D space. It is clear that plaintext images are clearly separated, while the scrambled ones are mixed together.}
	\label{fig:2Ddist}
\end{figure}

Next, we use PCA to reduce the dimensionality of the classification vectors of original and scrambled images. We randomly chose two labels from ImageNet, each consisting of 50 samples. We present our generated 2D representations in Figure \ref{fig:2Ddist}. It is clear that while the classification vectors for the two original groups of images are clearly separated, all the scrambled images are grouped together. Additionally, the scrambled images' representation is clearly separated from both groups of original images. These results further show that our scrambling process is effective in obfuscating the original label of the image.


\begin{table}
	\begin{center}
		\begin{tabular}{|c|c|} 
			\hline
			Top-1 Mean Intersection Rate & 0.07\% \\
			\hline
			Top-5 Mean Intersection Rate & 0.52\%\\
			\hline
			Top-1 Mean Reduction & 76\%\\
			\hline
			Top-5 Mean Reduction & 87\%\\
			\hline
		\end{tabular}
		\caption{Comparison of the label intersection and relative confidence scores for plaintext and scrambled images.}
		\label{tab:predictions-vectors-results}
	\end{center}
\end{table}

\subsection{Reconstruction of Scrambled Images}
\label{subsec:scramblingReconstruct}
In previous sections we showed that our scrambled images are not discernible by human users, and that they induce a higher degree of entropy in their classification. Our goal now is to determine whether our scrambled images are more difficult to reconstruct. We used an autoencoder in two experimental settings: \textit{a)} receive the original image as input, and reconstruct it; \textit{b)} receive the scrambled image as input, and reconstruct the original image.

As our autoencoder we used the well-known DCGAN architecture \cite{radford2015unsupervised}. We used DCGAN's discriminator, which we augmented with additional residual blocks, as our encoder. DCGAN's generator was used as our decoder. For \MethodName's setup we used the InceptionV3 architecture as our cloud-based model, and ResNet50 as our Encoder. 

For this experiment we randomly selected 100 labels from ImageNet, and then retrieved all 150 images associated with each label. We therefore created a set of 15,000 images. We than randomly split these 15,000 images into train and test sets, using a 90\%/10\% split. This process was repeated twice, and we present their averaged results in Table \ref{tab:AETrainMse2}. To enable faster convergence (i.e., using less training samples), all images were reduced to a quarter of their original size (128x128 instead of 256x256). 

\begin{figure}
	\begin{subfigure}{.23\textwidth}
		\centering
		\includegraphics[width=\textwidth]{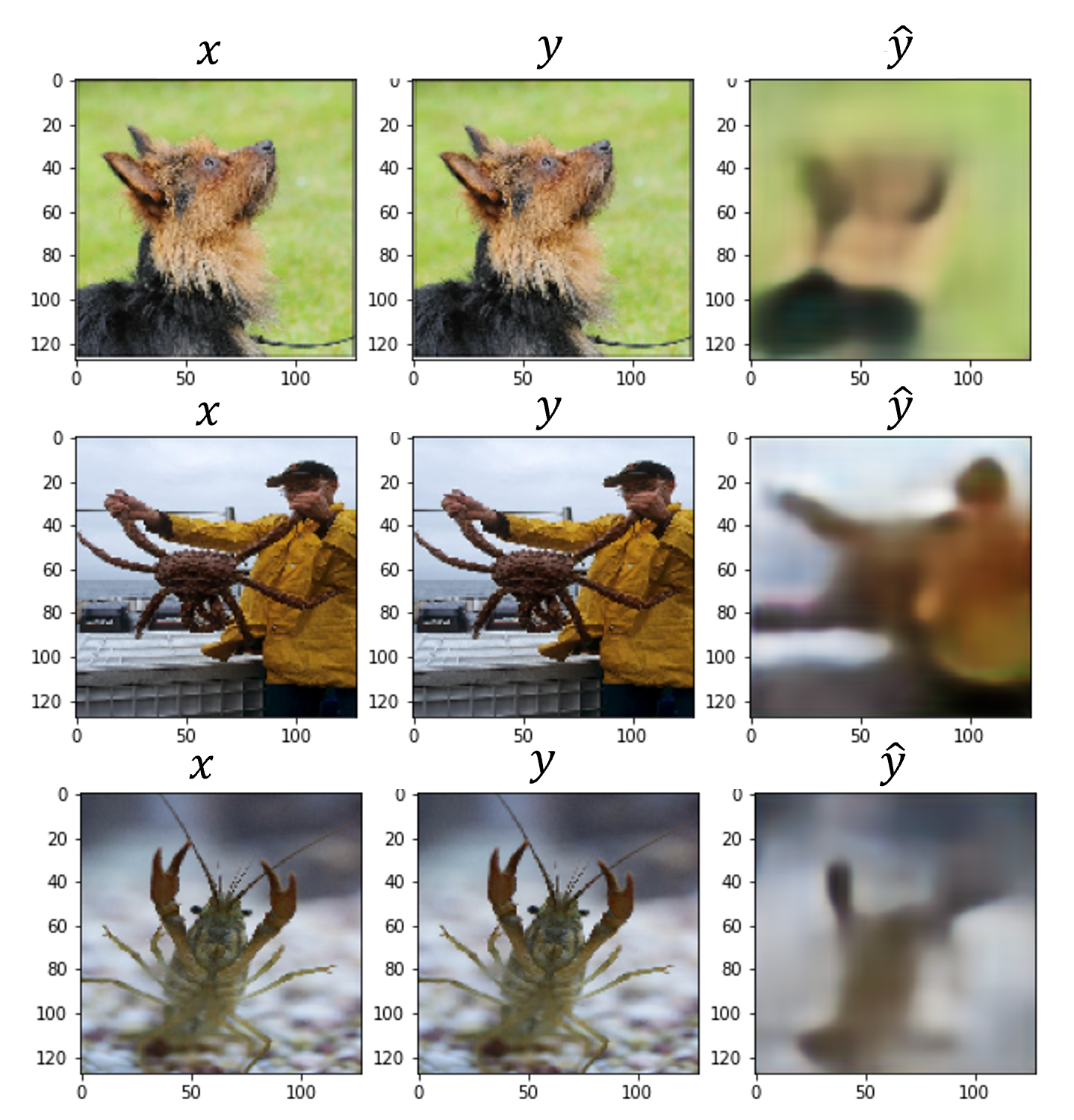}
		\caption{}
		\label{fig:70otest}
	\end{subfigure}
	\begin{subfigure}{.23\textwidth}
		\centering
		\includegraphics[width=\textwidth]{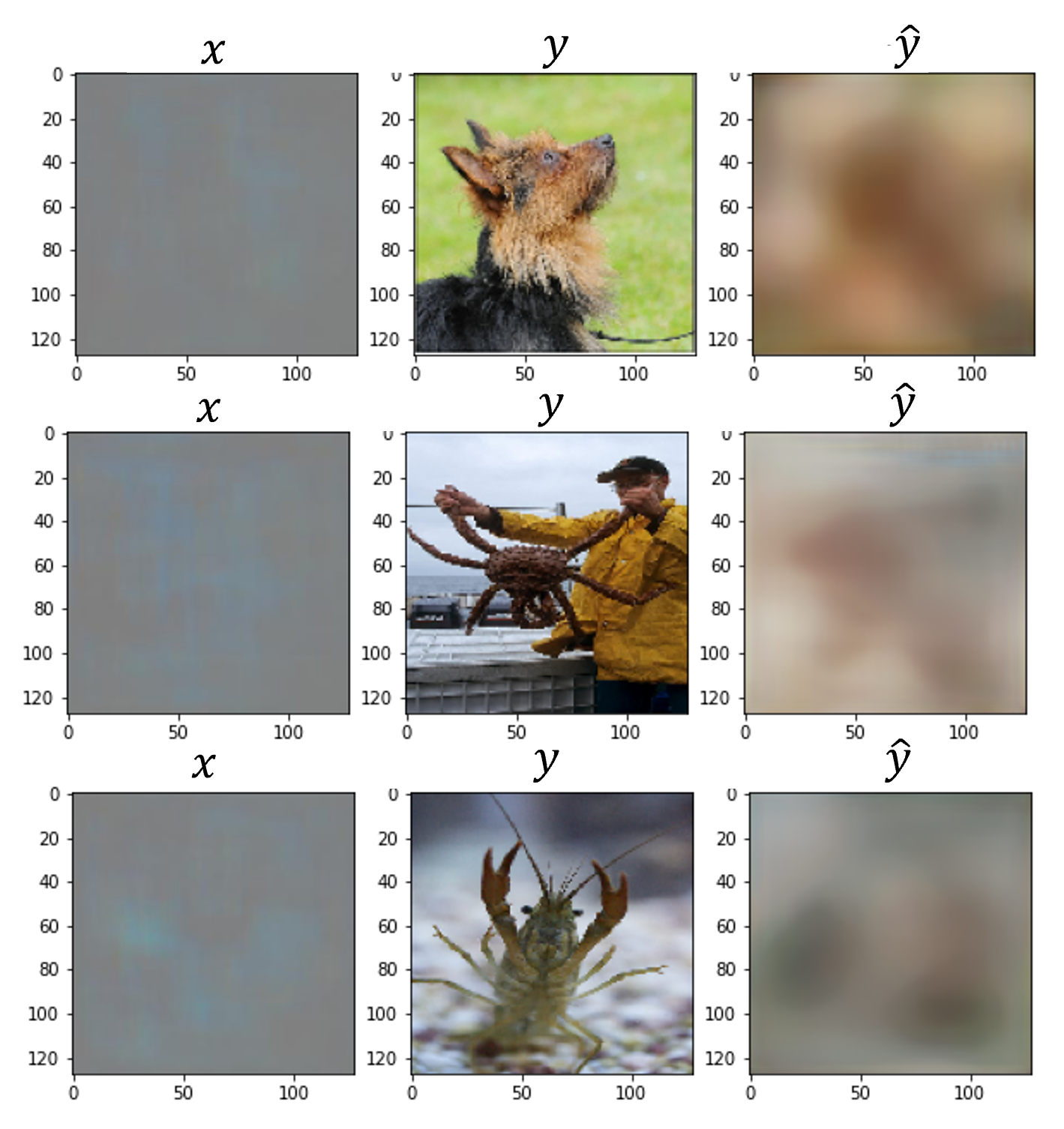}
		\caption{}
		\label{fig:70gtest}
	\end{subfigure}
	\caption{The inputs ($x$), targets ($y$) and reconstructions ($\hat{y}$) of images. Figure \ref{fig:70otest} presents the use-case where we receive the plaintext image as input, and \ref{fig:70gtest} presents the use-case where we receive the scrambled image.}

	\label{fig:autoencoder training}
\end{figure}

We define difficulty to reconstruct as the \textit{number of samples needed for the training of autoencoder}. This definition is crucial for Section \ref{subsec:AnalyzingRobustness}, where we analyze our model's robustness. The results of our evaluation are presented in Table \ref{tab:AETrainMse2} and Figure \ref{fig:70gtest}. While this training set size is sufficient for a decent reconstruction of images when applied on the original images, reconstruction quality for scrambled images is so low that the images are incomprehensible.

\begin{table}
	\footnotesize
	\centering
	\begin{tabular}{|c|c|c|} 
		\hline
		Images Type & Test Loss (MSE) & $\sigma$  \\
		\hline
		Original & 0.021 & 0.004\\
		\hline
		Scrambled & 0.061 & 0.001\\
		\hline
	\end{tabular}
	\caption{The autoencoder's reconstruction loss when applied on original and scrambled images}
	\label{tab:AETrainMse2}
\end{table}

\subsection{Analyzing the Robustness of Our Approach}
\label{subsec:AnalyzingRobustness}

As explained in Section \ref{sec:proposedMethod}, only the encrypted images ever leave the organizational network. As a result, attackers are likely to have to resort to a practically infeasible brute-force search to discover our randomly-set Generator weights. However, we are unable at this time to provide a theoretical proof to this claim, and therefore provide an \textit{empirical proof}. We do so by creating a scenario in which the adversary has access to \textit{additional information that enables a more effective attack against our approach}. We then use the amount of information required to carry out this attack as a bound on the number of images that can be securely scrambled by any one given key.\\

\noindent \textbf{The proposed attack scenario.} Assume that an adversary has gained access to pairs of original-scrambled images (i.e., not only does the adversary have access to such sets, but she can also pair them accordingly). The adversary can now train a neural architecture---more specifically, an autoencoder---to reconstruct the original image from the encrypted one. This scenario, in fact, is exactly the setup described in the second experiments of Section \ref{subsec:scramblingReconstruct}, \textit{where we showed that 13,500 original/scrambled image pairs are insufficient for any meaningful reconstruction of scrambled images}. It is also very important to note that---as shown in use-case 5 in Section \ref{subsec:evaluationResults}--- the maximal number of scrambled images needed to train the IIN is 4,500 (for confidential data with 100 labels). 

The aforementioned statistics provide us with a bound on the number of images that can be safely scrambled by \MethodName. Given that 13,500 image pairs are not enough to mount a successful attack, and that 4,500 images \textit{at most} are needed to train our IIN, then we are able to safely use any given key for 9,000 submissions to the cloud. When this number is reached, all we have to do is re-initialize our key. Finally, we wish to emphasize the following: 
\begin{itemize}
	\item We stress again that the figure of 13,500 is a very loose bound, created using an unrealistic scenario that greatly favors the adversary. Moreover, the autoencoders were trained on images whose size was only a quarter of the original, thus making the reconstruction process easier.
	\item The process of generating a new key is instantaneous, and the training of the IIN required approximately five minutes on a laptop (see INN description in Section \ref{sec:proposedMethod}). Therefore, replacing the scrambling key has a negligible computational cost.
	\item This process can easily be made more secure with the deployment of multiple keys and IINs. Doing so will not only improve performance (see use-case 4 in Section \ref{subsec:evaluationResults}), but their gradual replacement will make an adversarial attack even more difficult.
\end{itemize}

\section{Conclusion}
We present \MethodName, a scrambling-based approach that offers many of the advantages of Homomorphic encryption at a fraction of the computational cost. We showed that when the number of labels in the confidential data is smaller than in the CMLS, our approach can achieve the same performance as the CMLS. For future work, we intend to conduct experiments on additional types of data, and explore the use of our approach in a multi-agent collaborative setting.


\end{document}